\documentclass{PoS}

\title{Backreaction in Axion Monodromy, 4-forms and the Swampland}

\ShortTitle{Backreaction in Axion Monodromy, 4-forms and the Swampland}

\author{\speaker{Irene Valenzuela}%
       \\
       Max-Planck-Institute for Physics, Munich\\
       E-mail: \email{ireneval@mpp.mpg.de}}

\abstract{Axion monodromy models can always be described in terms of an axion coupled to 3-form gauge fields with non-canonical kinetic terms. The presence of the saxions parametrising the kinetic metrics of the 3-form fields leads to backreaction effects in the inflationary dynamics. We review the case in which saxions backreact on the K\"ahler metric of the inflaton leading to a logarithmic scaling of the proper field distance at large field. This behaviour is universal in Type II string flux compactifications and consistent with a refinement of the Swampland Conjecture. The critical point at which this behaviour appears depends on the mass hierarchy between the inflaton and the saxions. However,  in tractable compactifications, such a hierarchy cannot be realised without leaving the regime of validity of the effective theory, disfavouring transplanckian excursions in string theory.}

\FullConference{Corfu Summer Institute 2016 "School and Workshops on Elementary Particle Physics and Gravity"\\
                 31 August - 23 September, 2016\\
                 Corfu, Greece}

\def\beq{\begin{equation}}
\def\eeq{\end{equation}}
\def\beqa{\begin{eqnarray}}
\def\eeqa{\end{eqnarray}}

\usepackage{cite}

\begin{document}

\section{Introduction}

Effective theories involving a transplanckian field range for a scalar have received a lot of attention in the recent years both from the phenomenology and the theory communities. The reason is twofold. First, they are important phenomenologically since a scalar field rolling down a potential for a transplanckian distance is the key feature of both large field inflation  and the cosmological relaxation proposal for the EW hierarchy problem \cite{Graham:2015cka}. And second, because of the apparent aversion that string theory shows to transplanckian axions.  Even if there is not any fundamental (only technical) problem with having a transplanckian axion from the point of view of quantum field theory, the embedding of these theories in string theory is full of control issues that always forces us to leave the regime of validity of the effective theory.   
Furthermore, recent constraints coming from the Weak Gravity Conjecture (WGC) \cite{ArkaniHamed:2006dz} have led people think that these technical difficulties are in fact a sign of an underlying fundamental obstruction for realising a transplanckian scalar field range in a consistent theory of quantum gravity. If this is the case, a trans-planckian field range might constitute one of the few quantitative criteria known to distinguish between the string landscape and the swampland\footnote{Quantum field theories which cannot be embedded in a consistent theory of quantum gravity (and therefore cannot be derived from a consistent string theory compactification) are said to belong to the Swampland \cite{Vafa:2005ui,Ooguri:2006in}.}. 

Here we will focus on the axion monodromy proposal \cite{Silverstein:2008sg,McAllister:2008hb}, which so far has been kept safe against quantum gravity arguments coming from the WGC constraining the axionic decay constants. Axion monodromy is a mechanism to induce a non-periodic potential (like a mass term) for an axion while preserving its discrete shift symmetry. This is achieved by inducing a multi-branched potential which converts the circle in field space into an helix. In this way, the field range can in principle be much bigger than the fundamental periodicity of the axion, while still enjoying the protection given by the shift symmetry against higher dimensional operators. The four-dimensional description of axion monodromy is given by coupling the axion to a Minkowski 4-form, which corresponds to the field strength of a 3-form gauge field living in the space-time dimensions. In the inflationary community this is known as the Kaloper-Sorbo description \cite{Kaloper:2008fb,Kaloper:2011jz} although it was first analysed in detail by Dvali \cite{Dvali:2005an,Dvali:2005zk} in the context of the strong CP problem.

As outlined, the embedding of these models in string theory leads to control issues when taking into account the backreaction of the inflaton into the other scalars of the compactification \cite{McAllister:2014mpa,Hebecker:2014kva,Blumenhagen:2014nba,Buchmuller:2014vda,Dudas:2015lga,Buchmuller:2015oma}. Therefore, the presence of a Kaloper-Sorbo coupling is not enough to guarantee a transplanckian field range. Here we will show how these backreaction issues appear in the four-dimensional description in terms of 4-forms, which in fact corresponds to a supersymmetric embedding of  the Kaloper-Sorbo Lagrangian in $N=1$ supergravity. The above backreaction issues can then be naturally described by considering field-dependent kinetic metrics for the 3-form fields, without need of invoking unknown higher dimensional operators. 

Backreaction effects can lead to interesting features on the inflationary potential and even reduce the field range. However, so far, these problems seemed to be model dependent and not necessarily linked to the Planck mass. A universal feature, though, is the backreaction of the saxions into the K\"ahler metric of the inflaton, leading to a logarithmic scaling of the proper field distance at large field. In \cite{Baume:2016psm} Palti and Baume emphasised this logarithmic scaling and related it to one of the Swampland Conjectures of Ooguri-Vafa in \cite{Ooguri:2006in}. The conjecture claims that there is always a tower of states whose mass decreases exponentially with the scalar vev for large values of the field. Therefore, the effective theory of a scalar is valid only up to a finite distance in field space. Based on the results obtained in Type IIA flux compactifications \cite{Baume:2016psm} and some hints coming from black hole physics \cite{Klaewer:2016kiy}, the authors of these two papers defined a Refined Swampland Conjecture, stating that the logarithmic scaling appears necessarily close after crossing the Planck scale. If this is true, it would be the first model-independent argument constraining the field range of axion monodromy to a sub-planckian value.

Our main motivation is to analyse in more detail these backreaction issues and discern if they necessarily appear around the Planck scale. We will recover the results of \cite{Baume:2016psm} regarding Type IIA flux compactifications by analysing the backreaction in the formulation in terms of 4-forms. Then we will show that the flux-independence of the critical field value at which the logarithmic scaling appears is a specific feature of these models and not a universal behaviour of string theory. The criterium is simple: everytime the inflaton potential can be set to zero without destabilising the rest of the scalars, the model enjoys the requirements to be able to delay the backreaction effects by generating a mass hierarchy between the inflaton and the other scalars \cite{Valenzuela:2016yny,Bielleman:2016olv}. However, a closer analysis reveals that, at least in flux toroidal compactifications, this mass hierarchy cannot be achieved without getting into trouble with the Kaluza-Klein scale and leaving the regime of validity of the effective theory \cite{Blumenhagen:2017cxt}, supporting this way the Refined Swampland Conjecure. 

This proceedings article is based on \cite{Valenzuela:2016yny} and the talk given at the ``Workshop on Geometry and
Physics'' at Ringberg Castle in Germany. It is organised as follows. In section \ref{sec:4forms} we review the Kaloper-Sorbo description and clarify the differences when embedded in a supergravity formulation. In section \ref{sec:log} we discuss the backreaction on the K\"ahler metric of the inflaton leading to the aforementioned logarithmic scaling. We then analyse this effect in section \ref{sec:string} in Type IIA flux compactifications and models including D7-brane position moduli in Type IIB. The latter are promising to provide a mass hierarchy that might delay backreaction effects in field space. We leave section \ref{sec:con} for conclusions.

\section{Description in terms of Minkowski 4-forms\label{sec:4forms}}

Every four-dimensional scalar potential for an axion can always be rewritten in terms of a coupling to a 4-form field-strength  $F_4$. The simplest case of a mass term can be obtained from the following Lagrangian \cite{Dvali:2005an,Dvali:2005zk} (see also \cite{Kaloper:2008fb,Kaloper:2011jz,Marchesano:2014mla,Dudas:2014pva}),
\beq
\mathcal{L}=-f^2(d\phi)^2- F_4\wedge *F_4+2mF_4\phi
\label{KS}
\eeq
where $f$ is the decay constant of the axion $\phi$. 
Upon integrating out the Minkowski 3-form field, one gets a quadratic scalar potential 
\beq
*F_4=f_0+m\phi\equiv \rho(\phi)\rightarrow V=(f_0+m\phi)^2
\eeq
with different branches labelled by $f_0$, a possible constant value of the 4-form field strength in the vacuum. The above scalar potential is indeed invariant under the combined discrete shift 
\beq
f_0\rightarrow f_0 + c\ ,\quad \phi\rightarrow \phi -c/m
\label{shift}
\eeq
which identifies gauge equivalent branches when $c/m=2\pi f$. Tunneling between the different branches is mediated by nucleating a membrane electrically charged under the 3-form field, but the tunneling rate is not fast enough to constrain large field inflation significantly \cite{Ibanez:2015fcv,Hebecker:2015zss,Brown:2016nqt}. Invariance under the gauge symmetry of the 3-form field also constrains the form of the higher dimensional corrections, which should appear as functions of the field strength $F_4$ or $\rho(\phi)$ over the cut-off scale \cite{Kaloper:2008fb,Kaloper:2011jz,Kaloper:2014zba}. This guarantees in turn that the shift symmetry (\ref{shift}) is also preserved. Upon integrating out the 3-form field, corrections depending on $F_4$ will give rise to corrections depending on the scalar potential itself,
\beq
\delta\mathcal{L}=\sum_n c_n\left(\frac{F_4}{M_p^4}\right)^n\rightarrow \delta V =\sum_n c_n \left(\frac{\rho(\phi)}{M^4_p}\right)^n =\sum_n c_n\left(\frac{V_0}{M^4_p}\right)^n 
\label{corrections}
\eeq
so that they remain under control as long as the potential energy is subplanckian, even if $\phi$ takes transplanckian values. Finally, corrections going as powers of $F_4\rho(\phi)$ will give rise to higher derivative corrections for the axion \cite{Kaloper:2016fbr}, suppressed during slow-roll inflation.

The generalisation to multiple axions enjoying more elaborated scalar potentials is given by
\beq
\mathcal{L}=K_{ij}(s,\phi)d\phi^id\phi^j+K_{ij}(s,\phi)ds^ids^j-Z_{ab}(s)F_4^a\wedge *F_4^b +2F_4^a\rho_a(\phi)+V_{loc}(s)
\label{KSgeneral}
\eeq
where we have included the possible presence of non-axionic scalars (saxions $s^i$). Since the saxions are not periodic, their scalar potential can receive contributions that cannot be written in terms of 4-form couplings and which we denote as $V_{loc}(s^i)$. They will also appear parametrising the kinetic metric of the 3-form fields, which is indeed related to the K\"ahler metric of the scalar manifold if there is $N=1$ supersymmetry. Let us remark that all the dependence of the axions in the scalar potential only appears within the shift invariant functions $\rho_a(\phi)$ which couple to the 4-forms\footnote{One can also have periodic corrections that are shift invariant by themselves, like cosine-like terms. However, also these terms can be written in terms of a composite or effective 4-form with non-canonical kinetic term \cite{Dvali:2005an,Garcia-Valdecasas:2016voz}.}. Upon integrating out the 3-form fields, one gets
\beq
\label{pot}
V=Z_{ab}(s)^{-1}\rho_a(\phi)\rho_b(\phi)+V_{loc}(s)\ .
\eeq
In \cite{Bielleman:2015ina} it was checked that the full four-dimensional closed string scalar potential of Type IIA/B flux orientifold compactifications on a Calabi-Yau three-fold can be written as (\ref{KSgeneral}), where the 4-forms arise from dimensionally reducing the higher RR and NS p-form fields. The explicit form of $Z(s)$ and $\rho(\phi)$ (which we display in section \ref{sec:IIA}) was obtained by dimensionally reducing the 10d action and keeping track of all the 4-forms present in the compactification. A first analysis including open string fields has been performed in \cite{Carta:2016ynn}.

In the following we discuss the differences between the 4-form formulation (\ref{KSgeneral}) and the Kaloper-Sorbo description (\ref{KS}). These differences will be responsible for the control and backreaction issues which were not visible in (\ref{KS}).

\begin{itemize}
\item Non-linear couplings

The shift invariant functions $\rho(\phi)$ can include mixing and higher order terms for the axions, leading to more generic scalar potentials beyond the usual quadratic term.

\item Multiple 3-forms 

The presence of multiple 3-form fields imply that the higher dimensional corrections will not appear as powers of the scalar potential itself, but of the different $\rho$-functions,
\beq
\delta\mathcal{L}=\sum_n c_n\left(\frac{\prod_aF_4^a}{M_p^4}\right)^n\rightarrow \delta V =\sum_n c_n \left(\frac{\prod_a\rho_a(\phi)}{M^4_p}\right)^n\ .
\eeq
However, they are still under control as long as all $\rho(\phi)$ remain subplanckian.

\item Non-canonical field-dependent metrics

The mixing between axions and saxions in the scalar potential arises from the presence of the saxions in the kinetic metric of the 3-form fields, as can be seen in (\ref{pot}). This implies that, when displacing the inflaton from its minimum, the vacuum expectation values of the saxions will also change, backreacting on the effective theory of the inflaton. In particular, they can modify the field space metric leading to a reduction of the physical field range.

\end{itemize}

If one would be able to integrate out all the saxions, one would recover the Kaloper-Sorbo description only in terms of 4-forms and axions. The higher order corrections would then be proportional to $\rho(\phi)/m_s^4$ where the role of the cut-off scale is played by the saxion mass and not the Planck scale. However, as we will see, in many occasions the mass hierarchy between the inflaton and the saxion is not realisable in a consistent global compactification, invalidating such an effective theory. Therefore, we consider more appropriate to work in the 4-form formulation (\ref{KSgeneral}) where the saxions appear explicitly parametrising the kinetic metrics.

\section{Logarithmic scaling of the proper field distance\label{sec:log}}

In \cite{Ooguri:2006in} Ooguri and Vafa conjectured that the proper field range for a non-periodic scalar (saxion) scales logarithmically with the field at large field values. In \cite{Baume:2016psm,Klaewer:2016kiy} it was argued that the same behaviour applies to axions when taking into account the backreaction of the saxions. We explain the latter in the following.
The process of integrating out all the fields to obtain an effective theory involving only the inflaton is not feasible in general. In practice, one truncates the theory by freezing the saxions to their vevs at the minimum, and work with the remaining effective theory for the axion $\phi$ playing the role of the inflaton. However, this truncation is not valid when we displace $\phi$ away from the minimum, since the vevs of the saxions will generically depend on the inflaton vev as well. As a first approximation, one can then minimise the potential to obtain the leading order corrections of the saxions vevs as functions of $\phi$, and plug them again into the effective theory to analyse the backreaction effects.

From (\ref{pot}) we can see that the minimum of the saxions will be given by some function involving the $\rho(\phi)$ functions. According to the conjecture, they should be such that the saxion scales at least linearly with the inflaton at large field. If this is so, when plugging this result into the K\"ahler metric of the inflaton, the proper field distance of the axion will scale at best logarithmically for large field,
 \beq
\Delta\varphi=\int K_{\phi,\bar \phi}^{1/2} d\phi\sim \int \frac{1}{s(\phi)}d\phi\sim \int\frac{1}{\phi}d\phi\sim log(\phi)\ .
\eeq
Here we have used that $K=-log(S+S^*)$ where $S=s+i\phi$ for the sake of simplicity but it can be easily generalised to arbitrary K\"ahler potentials. Remarkably, we find this behaviour to appear universally in all F-term axion monodromy models constructed so far in the framework of flux string compactifications. Accompanying this logarithmic scaling, there is a tower of massive states which become exponentially light, $M\sim s(\phi)^{-n}\sim e^{-n\Delta \varphi }$, and invalidates the effective theory, so that inflation cannot proceed in this regime. This leads to an upper bound on the physical field range available in F-term axion monodromy.

The important question is, therefore, when this logarithmic behaviour appears. Is it always linked to the Planck scale or can it be delayed far out in field distance in a model-dependent way? To answer the question, let us be more concrete and split the saxion vev as follows,
\beq
\langle s\rangle = s_0 +\delta s(\phi)
\label{saxion1}
\eeq
where $s_0$ is the value at the minimum, $\delta s(\phi)=\lambda\phi$ for large field and $\lambda$ is some proportional factor. The logarithmic behaviour will appear when $\delta s(\phi_c)\approx s_0$, ie. at $\phi_c=s_0/\lambda$. The critical canonical field distance before backreaction effects dominate is therefore given by
 \beq
\varphi_c=\int^{\phi_c} K_{\phi,\bar \phi}^{1/2} d\phi\sim \int^{\phi_c} \frac{1}{s}d\phi\sim \frac{\phi_c}{s_0}\sim \frac{1}{\lambda}
\eeq
In the string theory examples studied in \cite{Baume:2016psm}, $\phi_c$ and $s_0$ scaled in the same way with the fluxes, leading to $\lambda\sim \mathcal{O}(1)$ in Planck units. This implied that the logarithmic scaling of the proper field distance appears necessarily around the Planck scale.

However, this relation between $\phi_c$ and $s_0$ is an artefact of the models under consideration. The fact that, in those models, one cannot set the inflaton mass to zero without destabilising the saxion makes impossible to suppress the backreaction. A more promising avenue is to engineer a model with a flat direction, in which all scalars are stabilised but one, and then provide an additional source that stabilises the latter. If the limit of switching off this last source is well defined in the sense that we recover the old minimum with only one flat direction, then $\lambda$ will depend on the mass hierarchy between the inflaton and the saxions. This provides a fundamental and not only a technical reason to look for single field inflationary models or at least, models in which the inflationary trajectory is not or barely mixed with non-axionic fields. Whether this is achievable in string theory, it is not clear yet. To sum up, let us remark that in general we find that
\beq
\lambda\sim \left(\frac{m_\phi}{m_s}\right)^p
\eeq
with $p=0,1$ depending on the model. In the next section, we will analyse some string theory models exemplifying both possibilities.

\section{Backreaction issues in string models\label{sec:string}}
Let us now analyse these backreaction issues in concrete models of string theory. First, we will recover the results of \cite{Baume:2016psm} regarding Type IIA flux compactifications by using the formulation in terms of 4-forms. Then, we will propose some open string models which look promising to realise a flux-dependent $\lambda$, and therefore delay the backreaction effects to transplanckian values.

\subsection{Closed string sector of Type IIA\label{sec:IIA}}

The flux induced scalar potential of an orientifold Calabi-Yau compactification of Type IIA reads
\beq
V=\frac{e^{K_{cs}}}{s}\left[\frac{1}{2k}|\rho_0|^2+\frac{g_{ij}}{8k}\rho^i\rho^j+2kg_{ij}\tilde\rho^i\tilde\rho^j+k|\rho_m|^2+\frac{1}{k}c_{IJ}\rho_h^{I}\rho_h^{J}\right]+V_{loc}
\eeq
where we have already integrated out the 4-forms. More details of the derivation can be found in \cite{Bielleman:2015ina,Valenzuela:2016yny}. From here one can read the metrics $Z_{ab}(s)$ in (\ref{pot}) which indeed depend on the saxions via volume factors labelled by $k$ and the field metrics $g_{ij}$ and $c_{IJ}$ in the K\"ahler and complex structure moduli space respectively. The shift invariant $\rho(\phi)$ functions are given by
\beqa
\rho_0=e_0+ b^ie_i-\frac{m}{6}k_{i jk}b^ib^jb^k+k_{i jk}\frac12 q_ib^jb^k-h_0c_3^0- h_ic_3^i\nonumber\\
\rho_i=e_i+k_{ijk}b^jq^k-\frac{m}{2}k_{ijk}b^jb^k\nonumber\\
\tilde\rho_i=q_i-mb_i\nonumber\\
\rho_m=m
\label{rho}
\eeqa
and $\rho^I_h=h^I$ where $e_0,e_i,q_i,m,h_I$ are internal fluxes while $b^i,c_3^I$ are four dimensional axions coming from $B_2$ and $C_3$ respectively. Here $i=1,\dots,h^{1,1}$ and $I=1,\dots h^{2,1}+1$ zzz. Thanks to the separation between saxions and axions provided by the 4-forms, the minimisation procedure gets enormously simplified in this formalism. For simplicity, let us consider $h^{1,1}=h^{2,1}=1$. The vacuum expectation values of the saxions at the minimum of the potential read
\beqa
\label{vevs}
\quad s_0\sim \frac{\rho_{i,0}^{3/2}}{\rho_{h_0}\sqrt{\rho_m}}\ ,\quad
u_0\sim \frac{\rho_{i,0}^{3/2}}{\rho_{h_1}\sqrt{\rho_m}}\ ,\quad
t_0\sim \frac{\rho_{i,0}^{1/2}}{\sqrt{\rho_m}}
\eeqa
where $\rho_{i,0}=e_i+q^2/(2m)$ is the value of $\rho_i$ at the minimum. These values are modified away from the minimum in the following way.

Let us consider the inflaton to be the RR axion $\phi\equiv c_3^0$ (the same result can be obtained for $c_3^i$). It can be checked that for large field the dilaton depends linearly on $\rho_0$ which indeed is linear in $c_3^0$. Therefore,
 \beq
\varphi=\int K_{\phi\bar \phi}^{1/2} d\phi= \int \frac{1}{s(\phi)}d\phi\rightarrow \int \frac{1}{\phi}d\phi\simeq \ log(\phi)
\eeq
where we have used $K=-log(s)$. The logarithmic behaviour appears at the critical value $\phi_c$ satisfying $\delta s(\phi_c)\sim s_0$. This occurs when $\Delta\rho_0\gtrsim t_0\rho_{i,0}$ implying $\phi_c\sim \rho_{i,0}^{3/2}/(h_0 \sqrt{m})\sim s_0$. Therefore $\lambda\sim \mathcal{O}(1)$ and the proper field distance before the logarithmic scaling appears
\beq
\varphi_c=\int^{\phi_c} K_{\phi\bar \phi}^{1/2} d\phi\sim \frac{\phi_c}{s_0}\sim \lambda^{-1}\sim  \mathcal{O}(1)
\eeq
is flux independent and tied to the Planck scale. 

The same behaviour can be obtained if the inflaton is a NS axion $\phi\equiv b^i$. In that case, the saxion entering on the K\"ahler metric of the inflaton is the K\"ahler modulus $t$. The logarithmic behaviour will occur when $\Delta\rho_i\gtrsim \rho_{i,0}$ implying a critical value given by $b_c=\sqrt{\rho_{i,0}/m}\sim t_0$. Therefore, the proper field distance before backreaction effects dominate is again
 \beq
\varphi_c=\int^{\phi_c} K_{\phi\bar \phi}^{1/2} d\phi\sim \frac{b_c}{t_0}\sim \lambda^{-1}\sim  \mathcal{O}(1)
\eeq
flux independent and bounded by the Planck scale. The results do not change by including more K\"ahler or complex structure moduli.

The underlying reason for having $\lambda\sim   \mathcal{O}(1)$ gets manifest when computing the mass spectrum. The inflaton mass is proportional to $h_0$ for the the case of the RR axion or to $m$ in the case of the NS axion, and therefore cannot be set to zero without destabilising the corresponding saxions (the vevs in (\ref{vevs}) go to infinity in that limit).

\subsection{D7-brane position moduli}

In order to get a flux-dependent tunable $\lambda$ we need to engineer a model in which all scalars but one are stabilised, and identify this flat direction with the inflaton. Then we add an additional source that stabilises the flat direction in such a way that the new minimum can be understood as a deformation of the old minimum. If this is satisfied, then we can delay the critical field distance at which the backreaction becomes important without decreasing the K\"ahler metric of the inflaton (parametrised by the saxion vevs) in the same amount. The parameter $\lambda$  will then depend on the mass hierarchy between the inflaton and the saxions and  thus the proper field distance before the logarithmic scaling appears will not be necessarily linked to the Planck scale.

Good candidates to realise these features are models in which a D7-brane position modulus plays the role of the inflaton  \cite{Arends:2014qca,Hebecker:2014eua,Ibanez:2014kia,Ibanez:2014swa,Bielleman:2015lka,Bielleman:2016grv,Landete:2017amp}. In Type IIB toroidal orientifold compactifications the $N=1$ supergravity effective theory is given by
\beqa
K=-log((S+\bar S)(U_3+\bar U_3)-\frac12 (\Phi+ \bar\Phi)^2)-3log(T+T^*)-\sum_{i=1}^2log(U_i+U_i^*)\\
W=W_0(S,T,U_i)+\mu \Phi^2
\label{D7}
\eeqa
where $W_0(S,T,U_i)$ contains the necessary flux terms to stabilise all closed string moduli. Let us consider the inflaton to be parametrised by $\phi\equiv\Phi-\bar\Phi$. This effective theory contains two features which help to delay the backreaction effects:
\begin{itemize}
\item The saxions parametrising (at the minimum) the K\"ahler metric of the inflaton do not belong to the same $N=1$ supermultiplet than the inflaton.
\item The structure of the superpotential allows us to set the inflaton mass to zero by choosing $\mu=0$ without destabilising the rest of the scalars.
\end{itemize}
If one minimises the potential with respect to all scalars but keeping the inflaton free, one gets that for large field
\beq
K_{\phi\bar \phi}^{1/2}=\sqrt{su_3}\simeq \sqrt{s_0u_0}+\lambda\phi
\eeq
where $\lambda\sim m_\phi/m_s$ up to numerical factors and $m_s$ is the scale at which the dilaton and complex structure moduli are stabilised. Here $m_\phi$ is controlled by $\mu$ at leading order while $m_s$ by the fluxes entering in $W_0$ (see \cite{Bielleman:2016olv,Blumenhagen:2017cxt} for concrete examples). Therefore the critical proper field distance before the logarithmic scaling dominates 
 \beq
\varphi_c=\int^{\phi_c} K_{\phi\bar \phi}^{1/2} d\phi\sim \lambda^{-1}\sim\frac{m_s}{m_\phi}
\eeq
is flux dependent and in principle can be tuned larger than the Planck scale by generating a mass hierarchy between the inflaton and the saxions. This is a clear difference with respect to the previous models within the closed string sector of Type IIA, in which $\lambda$ was flux-independent.

The key question now is whether such a mass hierarchy can be realised in a full-fletched global compactification of string theory. Here we have two options, either decrease the inflaton mass by assuming $\mu\ll 1$ or increase the saxion masses by choosing large fluxes in $W_0$. The first option is not possible in the toroidal compactification under consideration, since $\mu$ is a flux integer and cannot be tuned smaller than one. Therefore, we are left with the second option. However, if we embed the model in a KKLT moduli stabilisation scenario, large fluxes get in tension with having an exponentially small $W_0$. Furthermore, one also has to ensure that backreaction effects do not destabilise the K\"ahler modulus, which would make the KKLT minimum disappear.  This in turn implies $\mu\ll W_0$ which is not possible for $\mu$ being a flux integer. Embedding the model in the Large Volume Scenario does not help, because the condition required to suppress backreaction becomes $\mu\ll 1/\mathcal{V}$, which is not possible for large volume $\mathcal{V}$ either \cite{Blumenhagen:2017cxt}. These difficulties can be traced back to the fact that we are trying to get the inflaton, which is stabilised at tree level by fluxes, lighter than the K\"ahler modulus, which is stabilised by non-perturbative effects. Therefore, a good alternative would be to consider an scenario in which all moduli are stabilised at tree level, which requires the presence of non-geometric fluxes in IIB or geometric fluxes in IIA. However, in that case, the required mass hierarchy cannot be realised either without getting the moduli masses heavier than the Kaluza-Klein scale \cite{Blumenhagen:2017cxt}. Therefore,  even if $\lambda$ is flux dependent and in principle tunable in these models, in practice such a tuning is not possible in the global compactification and the critical proper field distance $\varphi_c$ cannot be made larger than the Planck scale without leaving the regime where the employed effective supergravity theory is under control.
This supports the Refined Swampland Conjecture of \cite{Baume:2016psm,Klaewer:2016kiy}. 

Notice, however, that a possible loophole comes from the quantisation of $\mu$. If one can get an effective $\mu$ parameter depending not only on flux integers but vacuum expectation values of other fields, then in principle one might be able to get $\mu\ll 1$ and delay the backreaction. However, without a concrete realisation of this fine-tuning in string theory, it is difficult to argue one way or the other, and could very well be that we are naively sweeping these backreaction issues under the carpet by appealing to the almighty string landscape.

\section{Conclusions\label{sec:con}}

Axion monodromy models always admit a dual description in terms of coupling the axions to 4-form field strengths in four dimensions. This formulation is useful to make manifest the underlying shift symmetries of the system, study the tunneling transitions between different branches and constrain the form of the higher dimensional corrections. 
When embedded in $N=1$ supergravity, the kinetic metric of the 3-form fields will be parametrised by the saxions (non-periodic scalars) of the compactification, while all the dependence on the axions will appear only within the shift invariant functions $\rho(\phi)$, which couple linearly to the 4-forms. These field dependent metrics give rise to the backreaction issues obtained when realising axion monodromy in a concrete string theory model.

The question addressed in this article is whether there is any universal behaviour appearing when crossing the Planck scale that forbids transplanckian excursions in axion monodromy. Interestingly, as emphasised in \cite{Baume:2016psm} saxions backreact on the K\"ahler metric of the inflaton such that at large field the proper field distance scales logarithmically with the field itself. This logarithmic scaling is consistent with the behaviour predicted by one of the Swampland Conjectures in \cite{Ooguri:2006in}. We reproduce the results of \cite{Baume:2016psm}  regarding flux compactifications of Type IIA by analysing the backreaction effects directly in the formulation in terms of 4-forms. It turns out that the critical field distance before the logarithmic behaviour appears is flux-independent and bounded by the Planck scale.

We show, though, that it is possible to construct models in which this critical field distance is fiux-dependent and can be tuned larger than the Planck scale by realising a mass hierarchy between the inflaton and the saxions. This occurs whenever the inflaton mass can be set to zero without destabilising the rest of the scalars, so the true minimum is a deformation of the old minimum (the one obtained before stabilizing the inflaton). A good inflaton candidate to realise this feature is a D7-brane position modulus in Type IIB orientifold compactifications \cite{Bielleman:2016olv,Valenzuela:2016yny}.
However, the mass hierarchy required to delay backreaction is not achievable in a toroidal compactification of Type IIB (or in F-theory on $K3\times K3$) \cite{Blumenhagen:2017cxt}. It remains to be seen if it can be realised in a more elaborated Calabi-Yau compactification involving mixing terms with many other moduli, in such a way that one can appeal to fine-tuning arguments in the landscape to reduce the inflaton mass \cite{Hebecker:2014kva}. 

For the moment, the multiple failing attempts and the lack of a concrete model realising this mass hierarchy, support the Refined Swampland Conjecture \cite{Baume:2016psm,Klaewer:2016kiy}, for which every effective theory of a scalar field breaks down upon a transplanckian field excursion in a consistent theory of quantum gravity.
In the upcoming years, the clarification of this issue will be important in view of possible cosmological data, and will help deeping our understanding of the boundaries between the string landscape and the swampland.

\vskip2em
\noindent
\emph{Acknowledgments:} 
This article summarises a talk given at the ``Workshop on Geometry and
Physics'', which took place on November 20-25, 2016 at Ringberg Castle, Germany. The workshop
was dedicated to the memory of Ioannis Bakas.

\bibliography{references}  

\providecommand{\href}[2]{#2}\begingroup\raggedright\begin{thebibliography}{10}

\bibitem{Graham:2015cka}
P.~W. Graham, D.~E. Kaplan, and S.~Rajendran, ``{Cosmological Relaxation of the
  Electroweak Scale},'' {\em Phys. Rev. Lett.} {\bf 115} (2015), no.~22,
  221801,
\href{http://www.arXiv.org/abs/1504.07551}{{\tt 1504.07551}}.

\bibitem{ArkaniHamed:2006dz}
N.~Arkani-Hamed, L.~Motl, A.~Nicolis, and C.~Vafa, ``{The String landscape,
  black holes and gravity as the weakest force},'' {\em JHEP} {\bf 06} (2007)
  060,
\href{http://www.arXiv.org/abs/hep-th/0601001}{{\tt hep-th/0601001}}.

\bibitem{Vafa:2005ui}
C.~Vafa, ``{The String landscape and the swampland},''
\href{http://www.arXiv.org/abs/hep-th/0509212}{{\tt hep-th/0509212}}.

\bibitem{Ooguri:2006in}
H.~Ooguri and C.~Vafa, ``{On the Geometry of the String Landscape and the
  Swampland},'' {\em Nucl. Phys.} {\bf B766} (2007) 21--33,
\href{http://www.arXiv.org/abs/hep-th/0605264}{{\tt hep-th/0605264}}.

\bibitem{Silverstein:2008sg}
E.~Silverstein and A.~Westphal, ``{Monodromy in the CMB: Gravity Waves and
  String Inflation},'' {\em Phys.Rev.} {\bf D78} (2008) 106003,
\href{http://www.arXiv.org/abs/0803.3085}{{\tt 0803.3085}}.

\bibitem{McAllister:2008hb}
L.~McAllister, E.~Silverstein, and A.~Westphal, ``{Gravity Waves and Linear
  Inflation from Axion Monodromy},'' {\em Phys.Rev.} {\bf D82} (2010) 046003,
\href{http://www.arXiv.org/abs/0808.0706}{{\tt 0808.0706}}.

\bibitem{Kaloper:2008fb}
N.~Kaloper and L.~Sorbo, ``{A Natural Framework for Chaotic Inflation},'' {\em
  Phys.Rev.Lett.} {\bf 102} (2009) 121301,
\href{http://www.arXiv.org/abs/0811.1989}{{\tt 0811.1989}}.

\bibitem{Kaloper:2011jz}
N.~Kaloper, A.~Lawrence, and L.~Sorbo, ``{An Ignoble Approach to Large Field
  Inflation},'' {\em JCAP} {\bf 1103} (2011) 023,
\href{http://www.arXiv.org/abs/1101.0026}{{\tt 1101.0026}}.

\bibitem{Dvali:2005an}
G.~Dvali, ``{Three-form gauging of axion symmetries and gravity},''
\href{http://www.arXiv.org/abs/hep-th/0507215}{{\tt hep-th/0507215}}.

\bibitem{Dvali:2005zk}
G.~Dvali, ``{A Vacuum accumulation solution to the strong CP problem},'' {\em
  Phys. Rev.} {\bf D74} (2006) 025019,
\href{http://www.arXiv.org/abs/hep-th/0510053}{{\tt hep-th/0510053}}.

\bibitem{McAllister:2014mpa}
L.~McAllister, E.~Silverstein, A.~Westphal, and T.~Wrase, ``{The Powers of
  Monodromy},'' {\em JHEP} {\bf 09} (2014) 123,
\href{http://www.arXiv.org/abs/1405.3652}{{\tt 1405.3652}}.

\bibitem{Hebecker:2014kva}
A.~Hebecker, P.~Mangat, F.~Rompineve, and L.~T. Witkowski, ``{Tuning and
  Backreaction in F-term Axion Monodromy Inflation},'' {\em Nucl. Phys.} {\bf
  B894} (2015) 456--495,
\href{http://www.arXiv.org/abs/1411.2032}{{\tt 1411.2032}}.

\bibitem{Blumenhagen:2014nba}
R.~Blumenhagen, D.~Herschmann, and E.~Plauschinn, ``{The Challenge of Realizing
  F-term Axion Monodromy Inflation in String Theory},'' {\em JHEP} {\bf 01}
  (2015) 007,
\href{http://www.arXiv.org/abs/1409.7075}{{\tt 1409.7075}}.

\bibitem{Buchmuller:2014vda}
W.~Buchmuller, C.~Wieck, and M.~W. Winkler, ``{Supersymmetric Moduli
  Stabilization and High-Scale Inflation},'' {\em Phys. Lett.} {\bf B736}
  (2014) 237--240,
\href{http://www.arXiv.org/abs/1404.2275}{{\tt 1404.2275}}.

\bibitem{Dudas:2015lga}
E.~Dudas and C.~Wieck, ``{Moduli backreaction and supersymmetry breaking in
  string-inspired inflation models},'' {\em JHEP} {\bf 10} (2015) 062,
\href{http://www.arXiv.org/abs/1506.01253}{{\tt 1506.01253}}.

\bibitem{Buchmuller:2015oma}
W.~Buchm\"uller, E.~Dudas, L.~Heurtier, A.~Westphal, C.~Wieck, and M.~W.
  Winkler, ``{Challenges for Large-Field Inflation and Moduli Stabilization},''
  {\em JHEP} {\bf 04} (2015) 058,
\href{http://www.arXiv.org/abs/1501.05812}{{\tt 1501.05812}}.

\bibitem{Baume:2016psm}
F.~Baume and E.~Palti, ``{Backreacted Axion Field Ranges in String Theory},''
  {\em JHEP} {\bf 08} (2016) 043,
\href{http://www.arXiv.org/abs/1602.06517}{{\tt 1602.06517}}.

\bibitem{Klaewer:2016kiy}
D.~Klaewer and E.~Palti, ``{Super-Planckian Spatial Field Variations and
  Quantum Gravity},'' {\em JHEP} {\bf 01} (2017) 088,
\href{http://www.arXiv.org/abs/1610.00010}{{\tt 1610.00010}}.

\bibitem{Valenzuela:2016yny}
I.~Valenzuela, ``{Backreaction Issues in Axion Monodromy and Minkowski
  4-forms},''
\href{http://www.arXiv.org/abs/1611.00394}{{\tt 1611.00394}}.

\bibitem{Bielleman:2016olv}
S.~Bielleman, L.~E. Ib\'a\~nez, F.~G. Pedro, I.~Valenzuela, and C.~Wieck,
  ``{Higgs-otic Inflation and Moduli Stabilization},'' {\em JHEP} {\bf 02}
  (2017) 073,
\href{http://www.arXiv.org/abs/1611.07084}{{\tt 1611.07084}}.

\bibitem{Blumenhagen:2017cxt}
R.~Blumenhagen, I.~Valenzuela, and F.~Wolf, ``{The Swampland Conjecture and
  F-term Axion Monodromy Inflation},''
\href{http://www.arXiv.org/abs/1703.05776}{{\tt 1703.05776}}.

\bibitem{Marchesano:2014mla}
F.~Marchesano, G.~Shiu, and A.~M. Uranga, ``{F-term Axion Monodromy
  Inflation},'' {\em JHEP} {\bf 09} (2014) 184,
\href{http://www.arXiv.org/abs/1404.3040}{{\tt 1404.3040}}.

\bibitem{Dudas:2014pva}
E.~Dudas, ``{Three-form multiplet and Inflation},'' {\em JHEP} {\bf 12} (2014)
  014,
\href{http://www.arXiv.org/abs/1407.5688}{{\tt 1407.5688}}.

\bibitem{Ibanez:2015fcv}
L.~E. Ib\'a\~nez, M.~Montero, A.~Uranga, and I.~Valenzuela, ``{Relaxion
  Monodromy and the Weak Gravity Conjecture},'' {\em JHEP} {\bf 04} (2016) 020,
\href{http://www.arXiv.org/abs/1512.00025}{{\tt 1512.00025}}.

\bibitem{Hebecker:2015zss}
A.~Hebecker, F.~Rompineve, and A.~Westphal, ``{Axion Monodromy and the Weak
  Gravity Conjecture},'' {\em JHEP} {\bf 04} (2016) 157,
\href{http://www.arXiv.org/abs/1512.03768}{{\tt 1512.03768}}.

\bibitem{Brown:2016nqt}
J.~Brown, W.~Cottrell, G.~Shiu, and P.~Soler, ``{Tunneling in Axion
  Monodromy},'' {\em JHEP} {\bf 10} (2016) 025,
\href{http://www.arXiv.org/abs/1607.00037}{{\tt 1607.00037}}.

\bibitem{Kaloper:2014zba}
N.~Kaloper and A.~Lawrence, ``{Natural chaotic inflation and ultraviolet
  sensitivity},'' {\em Phys. Rev.} {\bf D90} (2014), no.~2, 023506,
\href{http://www.arXiv.org/abs/1404.2912}{{\tt 1404.2912}}.

\bibitem{Kaloper:2016fbr}
N.~Kaloper and A.~Lawrence, ``{A Monodromy from London},''
\href{http://www.arXiv.org/abs/1607.06105}{{\tt 1607.06105}}.

\bibitem{Garcia-Valdecasas:2016voz}
E.~Garc{\'\i}a-Valdecasas and A.~Uranga, ``{On the 3-form formulation of axion
  potentials from D-brane instantons},''
\href{http://www.arXiv.org/abs/1605.08092}{{\tt 1605.08092}}.

\bibitem{Bielleman:2015ina}
S.~Bielleman, L.~E. Ib\'a\~nez, and I.~Valenzuela, ``{Minkowski 3-forms, Flux
  String Vacua, Axion Stability and Naturalness},'' {\em JHEP} {\bf 12} (2015)
  119,
\href{http://www.arXiv.org/abs/1507.06793}{{\tt 1507.06793}}.

\bibitem{Carta:2016ynn}
F.~Carta, F.~Marchesano, W.~Staessens, and G.~Zoccarato, ``{Open string
  multi-branched and K{\"a}hler potentials},'' {\em JHEP} {\bf 09} (2016) 062,
\href{http://www.arXiv.org/abs/1606.00508}{{\tt 1606.00508}}.

\bibitem{Arends:2014qca}
M.~Arends, A.~Hebecker, K.~Heimpel, S.~C. Kraus, D.~L\"ust, C.~Mayrhofer,
  C.~Schick, and T.~Weigand, ``{D7-Brane Moduli Space in Axion Monodromy and
  Fluxbrane Inflation},'' {\em Fortsch. Phys.} {\bf 62} (2014) 647--702,
\href{http://www.arXiv.org/abs/1405.0283}{{\tt 1405.0283}}.

\bibitem{Hebecker:2014eua}
A.~Hebecker, S.~C. Kraus, and L.~T. Witkowski, ``{D7-Brane Chaotic
  Inflation},'' {\em Phys. Lett.} {\bf B737} (2014) 16--22,
\href{http://www.arXiv.org/abs/1404.3711}{{\tt 1404.3711}}.

\bibitem{Ibanez:2014kia}
L.~E. Ib\'a\~nez and I.~Valenzuela, ``{The inflaton as an MSSM Higgs and open
  string modulus monodromy inflation},'' {\em Phys. Lett.} {\bf B736} (2014)
  226--230,
\href{http://www.arXiv.org/abs/1404.5235}{{\tt 1404.5235}}.

\bibitem{Ibanez:2014swa}
L.~E. Ib\'a\~nez, F.~Marchesano, and I.~Valenzuela, ``{Higgs-otic Inflation and
  String Theory},'' {\em JHEP} {\bf 01} (2015) 128,
\href{http://www.arXiv.org/abs/1411.5380}{{\tt 1411.5380}}.

\bibitem{Bielleman:2015lka}
S.~Bielleman, L.~E. Ib\'a\~nez, F.~G. Pedro, and I.~Valenzuela, ``{Multifield
  Dynamics in Higgs-otic Inflation},'' {\em JHEP} {\bf 01} (2016) 128,
\href{http://www.arXiv.org/abs/1505.00221}{{\tt 1505.00221}}.

\bibitem{Bielleman:2016grv}
S.~Bielleman, L.~E. Ib\'a\~nez, F.~G. Pedro, I.~Valenzuela, and C.~Wieck,
  ``{The DBI Action, Higher-derivative Supergravity, and Flattening Inflaton
  Potentials},'' {\em JHEP} {\bf 05} (2016) 095,
\href{http://www.arXiv.org/abs/1602.00699}{{\tt 1602.00699}}.

\bibitem{Landete:2017amp}
A.~Landete, F.~Marchesano, G.~Shiu, and G.~Zoccarato, ``{Flux Flattening in
  Axion Monodromy Inflation},''
\href{http://www.arXiv.org/abs/1703.09729}{{\tt 1703.09729}}.

\end{thebibliography}\endgroup
\bibliographystyle{utphys}


\end{document}